\documentclass{article}

\usepackage{arxiv}

\usepackage[utf8]{inputenc} 
\usepackage[T1]{fontenc}    
\usepackage{hyperref}       
\usepackage{url}            
\usepackage{booktabs}       
\usepackage{amsfonts}       
\usepackage{nicefrac}       
\usepackage{microtype}      
\usepackage{lipsum}		
\usepackage{graphicx}
\usepackage{natbib}
\usepackage{doi}
\usepackage{subcaption}
\usepackage{amsmath}

\title{When AI Blurs the Boundaries of Contribution: An Empirical Study of Authorship Calibration}

\author{ \href{https://orcid.org/0000-0003-1634-8856}{Célina TREUILLIER}\\
	Human-IST Institute, University of Fribourg\\
	Fribourg, SWITZERLAND\\
	\texttt{celina.treuillier@unifr.ch} \\
	\And
	\href{https://orcid.org/0000-0001-7834-0417}{Denis LALANNE} \\
	Human-IST Institute, University of Fribourg\\
	Fribourg, SWITZERLAND\\
	\texttt{denis.lalanne@unifr.ch} \\
}

\date{}

\renewcommand{\shorttitle}{When AI Blurs the Boundaries of Contribution}

\begin{document}
\maketitle

\begin{abstract}
	The broad adoption of Artificial Intelligence (AI), especially Generative AI, raises pressing questions about how users interact with these systems to produce new content. In this paper, we introduce the concept of authorship calibration, defined as users’ awareness of their actual authorship when interacting with AI. Using the CoAuthor dataset, we empirically examine how authorship calibration varies across users and how it relates to their frequency of AI use. Our results reveal high variability: users relying heavily on AI tend to misjudge their authorship, whereas those using AI less frequently exhibit more accurate authorship calibration. These findings suggest that AI can obscure users’ perception of their own authorship. In learning contexts, miscalibration can affect metacognitive monitoring and learning strategies, ultimately impacting learning outcomes. Fostering authorship calibration then appears essential for promoting responsible and educationally meaningful AI integration.
\end{abstract}


\section{Introduction}

The advent of sophisticated Artificial Intelligence (AI) algorithms, such as Large Language Models~(LLMs), has significantly impacted our daily habits. Among the various tasks for which such generative AI models can be employed, writing stands out as one of the most prominent, as LLMs can produce human‑like content rapidly and efficiently~\cite{gasaymeh2024university}. AI can then support writers by stimulating creativity, improving grammatical and orthographic accuracy, and enhancing the overall coherence and flow of a text~\cite{hwang2023exploring}. However, these capabilities also raise fundamental questions regarding ownership and authorship~\cite{draxler2024ai}, since parts of the output may be generated entirely by the AI without meaningful user contribution. More specifically, depending on how users interact with AI, the system can assume a range of roles, from simple spelling and grammar corrector to functioning as an undeclared ghostwriter~\cite{kleiman2024aiwriting}.

The education field emerges as a particularly significant area of adoption of generative AI~\cite{pelaez2024impact}, with learners now relying on such models to support them in achieving educational tasks. While providing them with permanent, personalized, and on‑demand assistance  ~\cite{adiguzel2023revolutionizing}, AI integration into educational contexts also introduces important challenges, especially concerning its effects on metacognitive processes~\cite{giannakos2025promise} that are essential for effective and autonomous learning~\cite{biggs1988role}. Among metacognitive processes, \textit{calibration}, or \textit{metacognitive calibration}, plays an important role:  it reflects the degree to which individuals' judgments about their understanding, capability, competence, or preparedness align with their actual demonstrated performance~\cite{lichtenstein1977calibration}. Given that AI usage in educational contexts may affect such metacognitive processes, it also can also potentially impact associated learning outcomes. 

In this paper, we aim to address both the challenges related to users’ authorship in AI‑assisted writing tasks and the associated impact on metacognitive processes, especially in the educational context. To guide our research work, we formulate the following Research Questions (RQs):
\begin{itemize}
    \item \textbf{RQ1}: How accurately do users evaluate their authorship when engaging in AI‑assisted tasks?   
    \item \textbf{RQ2}: Does the frequency of AI use influence authorship perception, and in what ways?
\end{itemize}

To investigate these questions, and building on the established notion of calibration~\cite{alexander2013calibration}, we introduce the concept of \textbf{authorship calibration}, defined as users’ accurate evaluation of their own contribution during an AI‑assisted task. High authorship calibration indicates that users are aware of how AI is used and integrated into the final output, while accurately evaluating their own authorship. In contrast, \textit{mis}calibration arises when users either overestimate or underestimate their actual contribution. In an educational context, as with traditional performance‑related \textit{mis}calibration, poor authorship calibration may hinder learning, potentially resulting in dissatisfaction, overconfidence, or reduced motivation.

We study authorship calibration in the context of writing tasks leveraging the CoAuthor dataset~\cite{lee2022coauthor}, and situates our findings within educational settings to discuss the possible implications for metacognitive processes and learning outcomes. This paper advances the field by (1) introducing the concept of authorship calibration in GAI environments, (2) offering insights into how GAI usage shapes this calibration, and (3) providing new evidence that accurate calibration can benefit educational contexts.

\section{Related Work}
\label{sec:sota}

\subsection{Generative AI and Education}
\label{sec:sota-GAI-educ}

The advent of AI and its spread to the general public has highly transformed our daily lives, and the education sector is highly impacted~\cite{pelaez2024impact}. Built upon AI algorithms that leverage existing content (text, audio, images, etc.), generative AI systems are able to automatically generate new content~\cite{epstein2023art}. Their integration in educational contexts presents promising opportunities for learners to access permanent, personalized, and on-demand feedback~\cite{adiguzel2023revolutionizing}, thereby potentially enhancing their overall learning experience~\cite{mogavi2023exploring}. AI is sometimes described as a transformative force capable of empowering learners and revolutionizing pedagogical practices~\cite{bahroun2023transforming}.  However, beside these potential benefits, AI in education also raises challenges, especially related to the development of cognitive processes underlying learning~\cite{giannakos2025promise, mittal2024comprehensive}. Of particular concern is the risk of reduced cognitive effort and learner engagement when AI is used for educational tasks~\cite{abdelghani2023generative} as learners can potentially become passive participants rather than active actors of their learning process.

To better understand how AI can impact learning, researchers have devoted considerable attention to identifying and describing user-AI interaction patterns. Gomez \textit{et al.}~\cite{gomez2025human} propose a taxonomy of seven human-AI collaboration patterns, differentiated by the temporal evolution of interactions and by which agent (human or AI) initiates the interaction. Complementing this framework, vanBerkel \textit{et al.}~\cite{van2021human} define three types of human-AI interaction based on the trigger mechanism for AI input: intermittent (explicit user request), continuous (implicit ongoing integration), and proactive (condition-dependent AI initiation). Within educational contexts, \textit{Memarian et al.}~\cite{memarian2024multidimensional} described a multidimensional taxonomy of learner-AI interaction centered on learning alignment and compatibility alignment, referring to the coherence among learning activities, learning experiences, and learning outcomes.

Beyond conceptualizing learner-AI interactions, a growing body of research examines how these interaction patterns impact learning performance. Specifically about AI-support for writing tasks, Nguyen \textit{et al.}~\cite{nguyen2024human} demonstrated that learners who engage more actively in the interaction process achieve higher performances. Similarly, Yang \textit{et al.}~\cite{yang2025modifying} revealed that active engagement in higher-order cognitive processes, such as critical evaluation, synthesis, and revision, is consistently associated with improved essay quality. Addressing this concern about reduced learner engagement, Arnold \textit{et al.}~\cite{arnold2021generative} demonstrated that by modifying the writer-AI interaction design, AI can remain supportive without replacing the writer. However, the role AI plays for writers can vary a lot: from editor supporting writers in the reviewing of their output, to co-author involving collaborative patterns, to ideas source providing inspiration and direction, to ghostwriter generating content that human authors may not transparently declare~\cite{kleiman2024aiwriting}. This raises fundamental questions about authorship and ownership in AI-assisted writing tasks~\cite{kim2025supporting}.

\subsection{Authorship in AI-assisted Writing}

The notion of authorship, traditionally referring to the state or fact of being the writer of a document or the creator of a work, becomes highly questionable when AI is integrated into the writing or creation process. As AI models now produce human-like text using powerful natural language models, considering AI as a co-author represents a legitimate conceptual question~\cite{bozkurt2024genai}. This raises complex issues regarding how to appropriately declare AI usage in a work, and what constitutes legitimate authorship in human-AI collaboration. Closely related is the concept of ownership, referring to the legal rights to control and profit from a work, which can be separate from its creator. Considering these related concepts of authorship and ownership, Draxler \textit{et al.}~\cite{draxler2024ai} introduced the "AI Ghostwriter Effect", describing situations where AI users do not attribute authorship to the AI, although they attribute ownership to it. Declaration of authorship by AI users then become skewed, as they avoid acknowledging the AI's contribution. 

These authorship and ownership concepts are particularly interesting in educational settings, where academic tasks may be jointly completed by learners and AI. This makes it highly difficult for teachers to evaluate knowledge acquisition and competence development as traditional assessment methods focus primarily on final products (i.e. product-based assessment) rather than the processes through which they were created (i.e. process-based assessment)~\cite{swiecki2022assessment}. Cheng \textit{et al.}~\cite{cheng2024evidence} develop a process-based assessment approach,  where the final product is not the unique object being evaluated; instead, the interaction between learners and AI becomes equally essential. Complementing this growing assessment process, technical approaches relying on natural language processing models are able to automatically classify the work as belonging to the learner or the AI~\cite{pan2025exploring, richburg2024automatic}, then informing about authorship and contribution in AI-assisted works.

Finally, an alternative approach focuses on making interaction patterns and text provenance visible~\cite{shibani2023visual, torres2019visualizing, hoque2024hallmark}. In educational contexts, this represents the opportunity to raise learners' awareness about their engagement in the learning process. For teachers, this gives important insights that can inform evaluation process or adaptations of teaching practices. This awareness-oriented approach connects authorship concerns directly to metacognitive processes, suggesting that learners understanding their own contribution can then adapt their cognitive mechanisms to foster a more effective learning.

\subsection{Metacognition, Self-Regulated Learning and Calibration}
\label{sec:sota-srl-cal}

In the educational field, metacognition refers to learners' awareness and regulation of their own cognitive processes, encompassing the planning, monitoring, and evaluating of their learning activities~\cite{flavell1979metacognition}. Metacognition plays a crucial role by enabling learners to become more strategic, reflective, and autonomous~\cite{biggs1988role}. Closely aligned with metacognition, Self-Regulated Learning (SRL) constitutes a well-established theoretical framework describing how learners actively control their own learning processes through a cyclical model involving planning, monitoring, and self-reflection phases~\cite{pintrich2004conceptual, zimmerman2002becoming}. SRL has been extensively studied in educational research, with empirical evidence consistently demonstrating that it serves as a robust predictor of academic success~\cite{khiat2022using}. Recently, Xu \textit{et al.}~\cite{xu2025enhancing} specifically examined the relationship between SRL and metacognition in AI environments, revealing that metacognitive support enhances SRL abilities, thereby improving the overall learning experience.

Within the broader metacognitive and SRL frameworks, the concept of \textit{calibration}, or \textit{metacognitive calibration}, represents an interesting construct~\cite{alexander2013calibration}. Calibration reflects the degree of correspondence between individuals' subjective judgments about their understanding, capability, competence, or preparedness and their objectively measured performance across these dimensions~\cite{lichtenstein1977calibration}. Well-calibrated learners demonstrate accurate awareness of what they know and what they do not know, enabling effective allocation of study resources and appropriate help-seeking behaviors. Conversely, miscalibrated learners exhibit poor metacognitive awareness and may be either overconfident, overestimating their competencies, or underconfident, underestimating their competencies. Calibration is related to SRL, with calibration accuracy influencing the efficiency of self-regulatory behaviors~\cite{stone2000exploring}. Well-calibrated learners can accurately adapt their learning strategies throughout the SRL cycle, thereby enhancing learning outcomes, whereas miscalibrated learners may implement ineffective strategies, resulting in suboptimal performance~\cite{alexander2013calibration}.

In today’s rapidly evolving educational landscape, ensuring that learners remain aware of their own abilities and of the role AI plays in their learning is a major challenge. Yet, to the best of our knowledge, metacognitive calibration remains poorly explored within the context of AI usage in education. In particular, studies exploring how calibration relates to authorship in AI‑assisted work are lacking. We therefore propose to investigate how learner–AI interactions influence what we refer to as authorship calibration, meaning a user’s awareness of his own contribution during an AI‑assisted task.

\section{The Concept of Authorship Calibration}
\label{sec:authorship_calibration}

In this work, we conceptualize calibration through the lens of authorship, introducing the concept of \textbf{authorship calibration}. This authorship calibration reflects the extent to which an individual's judgment about his contribution in an AI-assisted task aligns with his actual contribution in the final output. Authorship calibration is operationalized using Equation~\ref{eq:calibration}.

\begin{equation}
    \text{Authorship calibration} = \text{Declared Authorship} - \text{Actual Authorship} 
\label{eq:calibration}
\end{equation}

\begin{figure}[h!]
    \centering
    \includegraphics[width=0.5\linewidth]{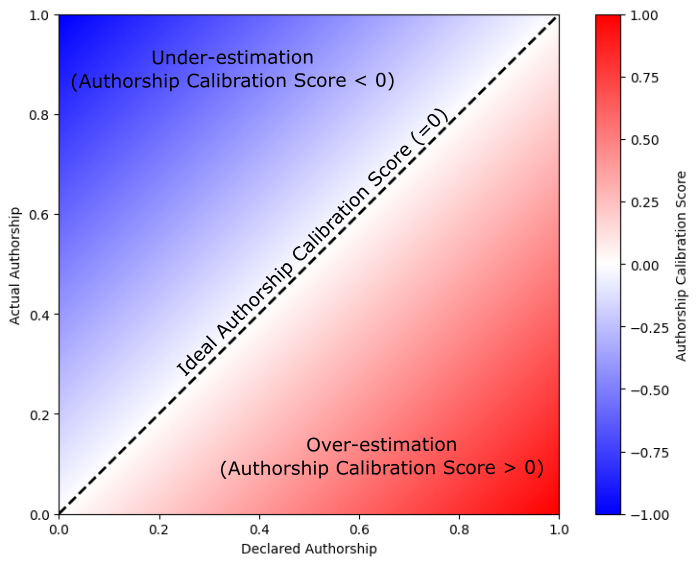}
    \caption{Variations in Authorship Calibration Scores as a Function of Declared and Actual Authorship}
    \label{fig:schema}
\end{figure}

Declared and actual authorship represent, respectively, the proportion of the final output that the user believes he has authored and the proportion he has actually authored. These proportion are expressed with continuous values ranging in $[0;1]$  (e.g., if the user authored half of the output ($50\%$), the corresponding value is $0.5$). The resulting authorship calibration score ranges in $[-1; 1]$, with a score of $0$ indicating a perfect calibration: the user’s perceived contribution matches his actual contribution. Negative scores reflect underestimation, meaning the user contributed more than declared; positive scores reflect overestimation, meaning the user contributed less than declared. The absolute magnitude value of the calibration score ($|\text{Authorship calibration}|$) indicates the degree of \textit{mis}calibration, with larger deviations from zero corresponding to poorer calibration. Figure \ref{fig:schema} illustrates how authorship calibration scores vary as a function of declared and actual authorship.

\section{Methodology}
\label{sec:methodo}

\subsection{Dataset}
\label{sec:dataset}

This study relies on the CoAuthor Dataset\footnote{\url{https://coauthor.stanford.edu/}}~\cite{lee2022coauthor}, which includes data about 1,445 AI-assisted writing sessions from 60 authors recruited on Amazon Mechanical Turk. The dataset encompasses two writing genres: 830 creative and 615 argumentative writing sessions.
Each writing session began with an initial prompt providing an overview of the assigned topic. Users could then choose to write independently (without using GPT) or request AI assistance through five GPT-generated suggestions, which they could ignore, modify, or incorporate without modifications. The dataset captures comprehensive interaction data, including GPT calls and keystroke events recording text insertions and deletions, cursor movements, and interactions with GPT suggestions.

Beyond interaction logs, the dataset provides additional resources including metadata about system configuration, and user survey responses. The metadata includes GPT parameters (temperature and frequency penalty settings) and some quantitative metrics characterizing GPT usage (i.e. number of queries, number of accepted suggestions, proportion of final text written by user, \textit{etc.}). The survey includes five sections assessing writer demographic and background information, perceived benefits of collaborative writing, user perceptions of GPT's capabilities and limitations, and overall writing experience. The CoAuthor Dataset is particularly well-suited to our research objectives, as it combines interaction data, essential for analyzing learner-AI collaboration patterns, with survey responses that enable assessment of authorship calibration.

\subsection{Users Categorization Based on the Frequency of AI Use}
\label{sec:users_categories}

To address our second research question (RQ2), we split users into two groups based on their AI usage. Following the methodology of Shibani et al.~\cite{shibani2023visual}, we classify users into Low- and High‑AI usage groups based on the median number of AI calls. This metric offers a comprehensive indicator of interaction intensity, capturing the full range of possible engagement with AI (e.g., requesting ideation suggestions, integrating suggestions with or without modification, \textit{etc.}). Users whose number of AI calls exceeds the median are assigned to the High‑AI usage group, while those below the median are assigned to the Low‑AI usage group. The full code is accessible on our github repository\footnote{\url{https://github.com/celinatreuillier/Authorship_Calibration_CoAuthor} (will be publicly available upon acceptance).}.

\subsection{Authorship Calibration Evaluation}
\label{sec:calibration-methodo}

To operationalize our introduced concept of authorship calibration, we leverage survey responses provided in the CoAuthor dataset. Importantly, participants were asked to answer the following question: "\textit{---\% of the essay/story is written by me (and the rest is written by taking the suggestions)}". Users then estimated the percentage ($[0-100\%]$) of their personal contribution in the final submitted text, as distinct from content originating from GPT suggestions. The actual percentage of text written by the human author is computed by comparing the number of user‑written sentences with the number of sentences originating from GPT suggestions. This percentage is provided in the CoAuthor metadata. We then compute authorship calibration using Equation \ref{eq:calibration}, which indicates the extent to which users accurately assess their authorship.

\section{Results}
\label{sec:results}

\subsection{Overall Authorship Calibration Scores}

From the original 1,445 writing sessions of CoAuthor, we kept only those with both interaction data and complete survey responses. Filtering sessions with missing data, 1,252 sessions remained for analysis (754 creative writing sessions and 508 argumentative writing sessions). For each session, authorship calibration is assessed by comparing declared authorship against actual authorship (See Section \ref{sec:authorship_calibration}). 

\begin{figure}[ht!]
    \centering
    \includegraphics[width=0.7\linewidth]{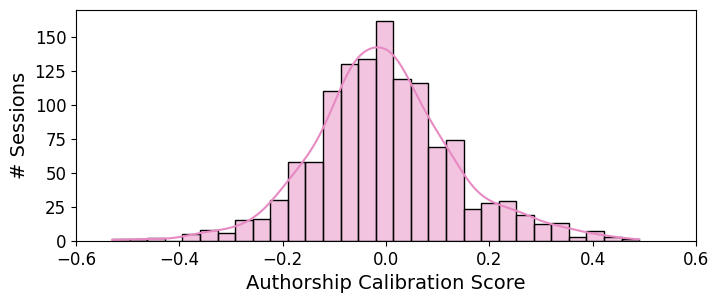}
    \caption{Distribution of Authorship Calibration Scores}
    \label{fig:histplot}
\end{figure}

The distribution of authorship calibration scores shows a roughly bell‑shaped pattern centered around zero (mean value $= -0.003$), indicating accurate calibration for most writing sessions (See Figure \ref{fig:histplot}). However, the Standard Deviation (SD) is high ($SD=0.138$), with scores ranging from $-0.53$ to $0.49$, indicating the presence of both strongly under‑ and over‑calibrated users. A limited skewness toward negative values is observed, suggesting a small tendency toward underestimation. Overall, these results highlight considerable variability, motivating further analysis of differences across Low- and High-AI usage groups.

\subsection{Authorship Calibration Across Low- and High-AI Usage Groups}

To deepen our analysis, we examine whether and how the frequency of AI use impact authorship calibration. As detailed in Section \ref{sec:users_categories}, we differentiate between Low-AI usage and High-AI usage comparing the number of calls made to AI (i.e. requests for AI-generated suggestions). In the filtered dataset of 1,251 writing sessions, the mean number of AI calls is  $12.98$ and varies greatly between users ($SD=9.6$), ranging from  $0$ calls to  $65$  calls. Relying on the median value of $11$  AI calls, we classified  647  sessions as High-AI usage and  605  sessions as Low-AI usage.

\begin{figure}[!ht]
\centering
\begin{subfigure}[t]{0.4\textwidth}
  \includegraphics[width=\linewidth]{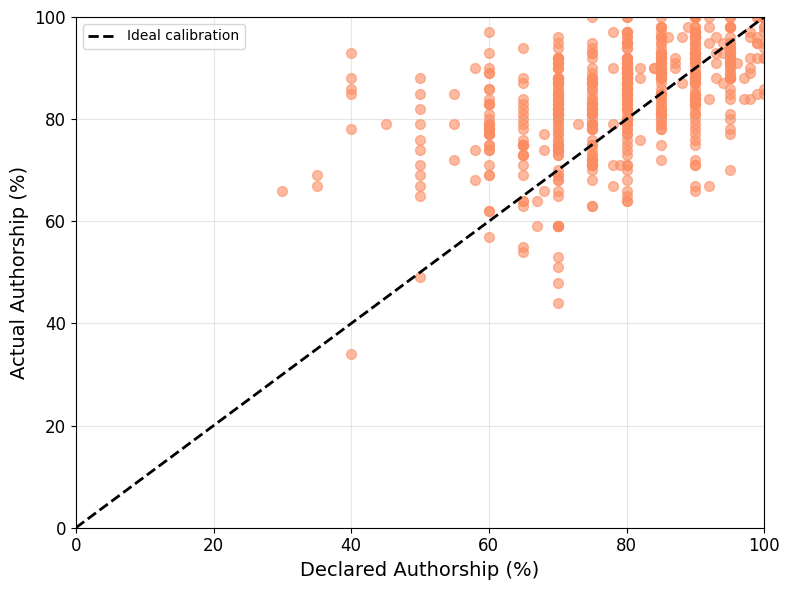}
  \caption{Low-AI usage}
  \label{fig:calibration_low}
\end{subfigure}
\begin{subfigure}[t]{0.4\textwidth}
  \includegraphics[width=\linewidth]{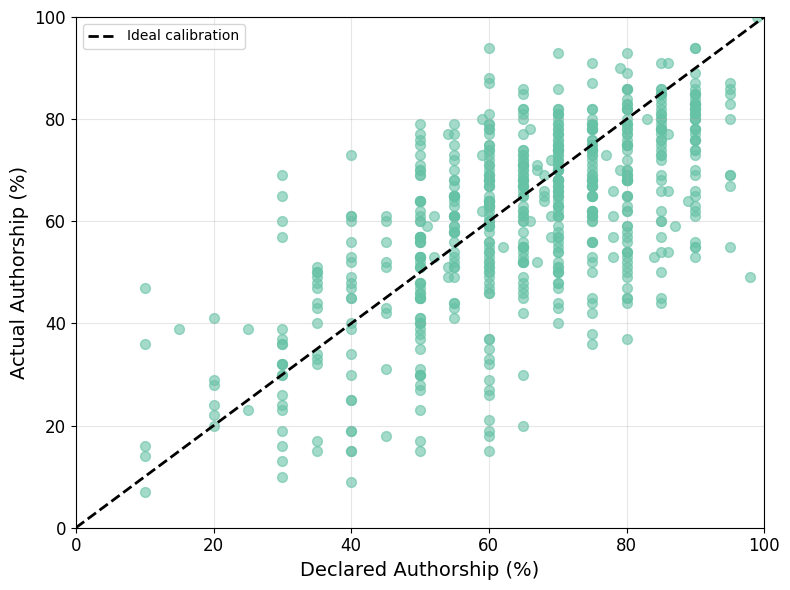}
  \caption{High-AI usage}
  \label{fig:calibration_high}
\end{subfigure}
\caption{Calibration curves.}
\label{fig:calibration_curves}
\end{figure}

\begin{figure}[!ht]
\centering
\begin{subfigure}[t]{0.45\textwidth}
  \includegraphics[width=\linewidth]{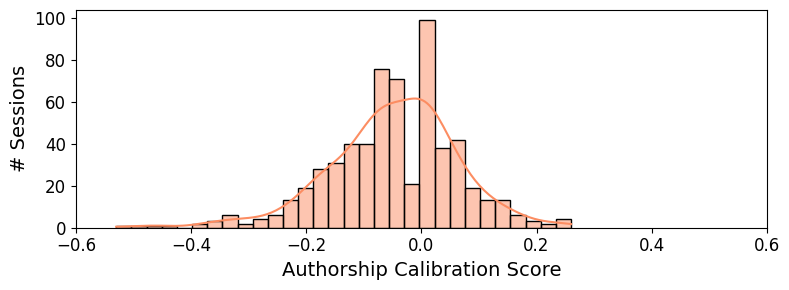}
  \caption{Low-AI usage}
  \label{fig:histplot_low}
\end{subfigure}
\begin{subfigure}[t]{0.45\textwidth}
  \includegraphics[width=\linewidth]{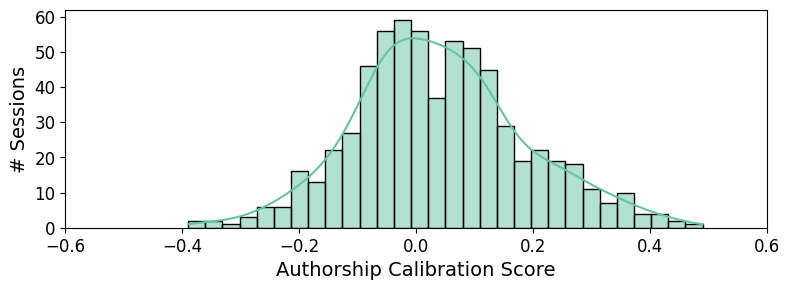}
  \caption{High-AI usage}
  \label{fig:histplot_high}
\end{subfigure}
\caption{Distribution of Authorship Calibration Scores in Users Groups.}
\label{fig:histplot_groups}
\end{figure}

\begin{figure}[!ht]
\centering
\begin{subfigure}[t]{0.4\textwidth}
  \includegraphics[width=\linewidth]{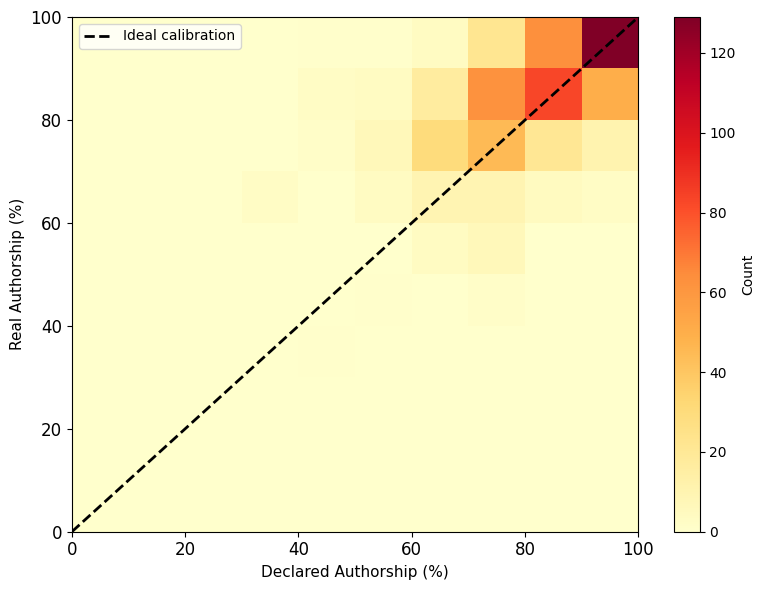}
  \caption{Low-AI usage}
  \label{fig:heatmap_low}
\end{subfigure}
\begin{subfigure}[t]{0.4\textwidth}
  \includegraphics[width=\linewidth]{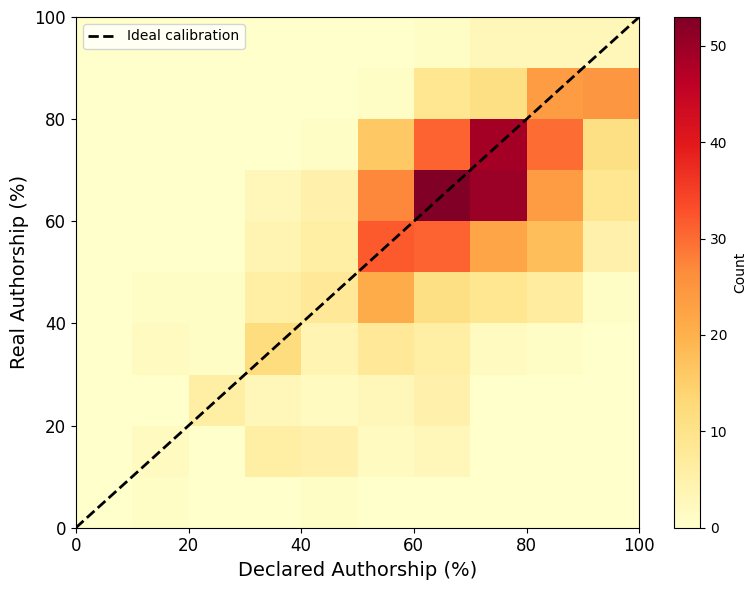}
  \caption{High-AI usage}
  \label{fig:heatmap_high}
\end{subfigure}
\caption{Heatmap of calibration curves.}
\label{fig:calibration_heatmaps}
\end{figure}

To compare calibration patterns across user groups, we plot calibration curves (Figure \ref{fig:calibration_curves}), where each point represents a user’s declared authorship positioned against the corresponding actual authorship, and the distribution of corresponding authorship calibration scores is presented in Figure \ref{fig:histplot_groups}. To further analyze how users are distributed within the calibration space, we also provide density heatmaps in Figure \ref{fig:calibration_heatmaps}, offering a more aggregated view of local concentrations and global patterns.

First, users from the the Low-AI usage group are predominantly concentrated in the upper-right quadrant, with the majority of both declared and real authorship values ranging between  $70\%-100\%$ (See Figure \ref{fig:calibration_low}). Low-GPT usage users are relatively evenly distributed on both sides of the ideal calibration line, though a slight underestimation tendency is observable (more user above the ideal calibration line). This is confirmed by the distribution of authorship calibration score that are skewed towards negative values (See Figure \ref{fig:histplot_low}), with a mean authorship calibration score of $-0.004$ ($SD=0.112$), ranging from  $-0.53$  to  $0.26$. This negative bias indicates that users with limited AI interaction during writing sessions tend to slightly underestimate their authorship, declaring authorship lower than the actual one. The corresponding heatmap confirms this pattern with the highest density in the  $90\%-100\%$  range for both declared and real authorship (See Figure \ref{fig:calibration_low}). This tight clustering around maximal authorship values demonstrates that writers with less frequent AI usage maintain high authorship, while mostly accurately evaluating it. 

Second, users from the High-AI usage group are broadly distributed in the calibration space, with both declared and actual authorship spanning from $10\%$ to almost $100\%$ (See Figure \ref{fig:calibration_high}). Specifically, there is considerable spread below the ideal calibration line (overestimation), particularly visible in the  $30\%-90\%$  declared authorship range. Corresponding authorship calibration values also show greater variability compared to the Low-GAI usage group, with a mean authorship calibration value of  $0.003$,  a higher standard deviation ($SD=0.146$) and a wider range of  $[-0.39,0.49]$ (See Figure \ref{fig:histplot_high}). This high variability suggests that users experience an inconsistent awareness of their actual authorship, while having a slight tendency to overestimate it. The heatmap distribution reinforces these observations, with the high densities ($20 - 50$  users) occurring across the  $50\%-90\%$  range for both declared and real authorship (See Figure \ref{fig:heatmap_high}). Importantly, a higher density appears below the ideal calibration line, particularly in regions where declared authorship of  $60\%-80\%$  corresponds to real authorship of only  $40\%-70\%$. Users relying more frequently on AI during the writing process then tend to overestimate their authorship. Finally, a statistical comparison further confirms that the distributions of authorship calibration scores differ significantly between Low‑AI and High‑AI users (Mann–Whitney U test, $p<0.05$).

\section{Discussion and Future Work}
\label{sec:limits}

Our experimental study reveals that while numerous users have accurate authorship calibration scores, an important proportion still show poor calibration, either overestimating or underestimating their authorship. This confirms that authorship calibration is not a trivial construct in AI‑assisted contexts, and that important differences persist across users, as not all of them are able to accurately assess it (RQ1). Besides, our results confirms that the accuracy of authorship calibration vary depending on the frequency of AI use during the writing task (RQ2). Specifically, users in the Low‑AI usage group show a more accurate authorship calibration, but a slight tendency to underestimate it. This may reflect stronger metacognitive awareness, as these users remain attentive to even minimal AI assistance and consequently minimize their own contribution. In contrast, users in the High‑AI usage group show poorer authorship calibration, with a tendency to overestimate it. Extensive AI interactions then appears to blur the distinction between human‑generated and AI‑generated content. This may be the results of an “effort blend” phenomenon: the cognitive work involved in prompting, selecting, and integrating AI suggestions is subjectively experienced as equivalent to generating original text, resulting in a higher sense of authorship.
These observations raise important concerns about authorship in AI‑assisted learning environments. Given the conceptual links between authorship calibration and higher‑order metacognitive processes, improving authorship calibration may represent a promising pathway to enhancing learning outcomes in AI‑assisted contexts.

Despite introducing the concept of authorship calibration and offering promising insights, this study has several limitations. First, authorship is examined only within writing tasks rather than in authentic educational settings. Exploring a wider range of educational tasks would allow for richer experiments, ultimately providing a more comprehensive understanding of learner–AI collaboration. Second, the distinction between Low‑ and High‑AI usage groups is based on a simple yet effective methodology. Applying more fine‑grained Learning Analytics pipelines to educational datasets would allow for richer classifications of learner–AI interaction patterns, ultimately providing richer insights into how specific interaction patterns influence authorship calibration.

Finally, this study opens promising directions for future work. A key avenue lies in designing and evaluating methods that support authorship calibration, e.g. through adapted user interfaces or personalized feedback. By increasing learners’ awareness about their actual contribution and fostering more accurate authorship calibration, such interventions may promote more appropriate metacognitive engagement, ultimately improving learning processes and outcomes. In addition, further research is needed to examine how authorship calibration relates to broader metacognitive processes, particularly within AI‑assisted learning environments.

\section{Conclusion}
\label{sec:ccl}

This paper introduces authorship calibration, the extent to which users accurately perceive their own authorship during AI-assisted tasks. Our results show clear variability across users, with authorship calibration being influenced by the level of AI usage. While some users maintain an accurate sense of authorship, others substantially misjudge their contribution, revealing how easily AI can blur boundaries between human- and AI-produced content.
Importantly, authorship calibration offers a powerful lens for understanding learning in AI‑assisted environments. Accurate authorship calibration may supports metacognitive monitoring and informed engagement, whereas miscalibration risks unproductive behaviors and weaker learning outcomes.  
As AI becomes broadly embedded in education settings, helping learners stay aware of their actual authorship appears essential. Designing tools and pedagogical strategies that strengthen authorship calibration will be key to fostering responsible, transparent, and educationally meaningful integration of AI in educational settings.

\bibliographystyle{unsrt}
\bibliography{references}

\end{document}